\documentclass{iau}

\usepackage{amsmath}
\usepackage{graphicx}
\usepackage{multirow}
\usepackage{xspace}
\usepackage{float}
\usepackage{color} 
\usepackage{ulem}
\usepackage{graphicx}	

\newcommand{\ie}{i.e.,\xspace}
\newcommand{\eg}{e.g.,\xspace}
                          
\newcommand{\msun}{\ifmmode \text{M}_{\odot} \else M$_{\odot}$\fi\xspace}

\newcommand{\Mstar}{\ifmmode M_{\star} \else $M_{\star}$\fi\xspace}
\newcommand{\Lstar}{\ifmmode L_{\star} \else $L_{\star}$\fi\xspace}

\newcommand{\Te}{$T_{\rm e}$\xspace}
\newcommand{\Ne}{$n_{\rm e}$\xspace}

\newcommand{\cmcub}{cm$^{-3}$}

\newcommand{\Ha}{\ifmmode \text{H}\alpha \else H$\alpha$\fi\xspace}
\newcommand{\Hb}{\ifmmode \text{H}\beta \else H$\beta$\fi\xspace}
\newcommand{\neiii}{\ifmmode [\text{Ne}\,\textsc{iii}] \else [Ne~{\scshape iii}]\fi\xspace}
\newcommand{\Neiii}{\ifmmode [\text{Ne}\,\textsc{iii}]\lambda 3869 \else [Ne~{\scshape iii}]$\lambda 3869$\fi\xspace}
\newcommand{\oii}{\ifmmode [\text{O}\,\textsc{ii}] \else [O~{\scshape ii}]\fi\xspace}
\newcommand{\Oii}{\ifmmode [\text{O}\,\textsc{ii}]\lambda 3726 + \lambda 3729 \else [O~{\scshape ii}]$\lambda 3726 + \lambda 3729$\fi\xspace}
\newcommand{\Oiiit}{\ifmmode [\text{O}\,\textsc{iii}]\lambda 4363 \else [O~{\scshape iii}]$\lambda 4363$\fi\xspace}
\newcommand{\hii}{\ifmmode \text{H}\,\textsc{ii} \else H~{\scshape ii}\fi\xspace}
\newcommand{\heii}{\ifmmode \text{He}\,\textsc{ii} \else He~{\scshape ii}\fi\xspace}
\newcommand{\Heii}{\ifmmode \text{He}\,\textsc{ii}\lambda 4686 \else He~{\scshape ii}$\lambda 4686$\fi\xspace}
\newcommand{\ariv}{\ifmmode [\text{Ar}\,\textsc{iv}] \else [Ar~{\scshape iv}]\fi\xspace}
\newcommand{\Ariv}{\ifmmode [\text{Ar}\,\textsc{iv}]\lambda 4711 + \lambda 4740 \else [Ar~{\scshape iv}]$\lambda 4711 + \lambda 4740$\fi\xspace}
\newcommand{\Niit}{\ifmmode [\text{N}\,\textsc{ii}]\lambda 5755 \else [N~{\scshape ii}]$\lambda 5755$\fi\xspace}
\newcommand{\hei}{\ifmmode \text{He}\,\textsc{i} \else He~{\scshape i}\fi\xspace}
\newcommand{\Hei}{\ifmmode \text{He}\,\textsc{i}\lambda 5876 \else He~{\scshape i}$\lambda 5876$\fi\xspace}
\newcommand{\nii}{\ifmmode [\text{N}\,\textsc{ii}] \else [N~{\scshape ii}]\fi\xspace}
\newcommand{\niis}{\ifmmode [\text{N}\,\textsc{ii}]_S \else [N~{\scshape ii}]$_S$\fi\xspace}
\newcommand{\Nii}{\ifmmode [\text{N}\,\textsc{ii}]\lambda 6584 \else [N~{\scshape ii}]$\lambda 6584$\fi\xspace}
\newcommand{\oiii}{\ifmmode [\text{O}\,\textsc{iii}] \else [O~{\scshape iii}]\fi\xspace}
\newcommand{\oiiis}{\ifmmode [\text{O}\,\textsc{iii}]_S \else [O~{\scshape iii}]$_S$\fi\xspace}
\newcommand{\Oiii}{\ifmmode [\text{O}\,\textsc{III}]\lambda 5007 \else [O~{\scshape III}]$\lambda 5007$\fi\xspace}
\newcommand{\sii}{\ifmmode [\text{S}\,\textsc{ii}] \else [S~{\scshape ii}]\fi\xspace}
\newcommand{\Sii}{\ifmmode [\text{S}\,\textsc{ii}]\lambda 6716 + \lambda 6731 \else [S~{\scshape ii}]$\lambda 6716 + \lambda 6731$\fi\xspace}
\newcommand{\ariii}{\ifmmode [\text{Ar}\,\textsc{iii}] \else [Ar~{\scshape iii}]\fi\xspace}
\newcommand{\Ariii}{\ifmmode [\text{Ar}\,\textsc{iii}]\lambda 7135 \else [Ar~{\scshape iii}]$\lambda 7135$\fi\xspace}
\newcommand{\siii}{\ifmmode [\text{S}\,\textsc{iii}] \else [S~{\scshape iii}]\fi\xspace}
\newcommand{\Siii}{\ifmmode [\text{S}\,\textsc{iii}]\lambda 9069 \else [S~{\scshape iii}]$\lambda 9069$\fi\xspace}

\newcommand{\rOii}{ \ifmmode [\text{O}\,\textsc{ii} ]\lambda 3726/3729 \else [O~{\scshape  ii}]$\lambda 3726/3729$\fi\xspace}
\newcommand{\rOiii}{\ifmmode [\text{O}\,\textsc{iii}]\lambda 4363/5007 \else [O~{\scshape iii}]$\lambda 4363/5007$\fi\xspace}
\newcommand{\rAriv}{\ifmmode [\text{Ar}\,\textsc{iv}]\lambda 4740/4711 \else [Ar~{\scshape iv}]$\lambda 4740/4711$\fi\xspace}
\newcommand{\rNii}{ \ifmmode [\text{N}\,\textsc{ii} ]\lambda 5755/6584 \else [N~{\scshape  ii}]$\lambda 5755/6584$\fi\xspace}
\newcommand{\rCliii}{\ifmmode [\text{Cl}\,\textsc{iii}]\lambda 5537/5517 \else [Cl~{\scshape iii}]$\lambda 5537/5517$\fi\xspace}
\newcommand{\rSiii}{\ifmmode [\text{S}\,\textsc{iii}]\lambda 6312/9532 \else [S~{\scshape iii}]$\lambda 6312/9532$\fi\xspace}
\newcommand{\rSii}{ \ifmmode [\text{S}\,\textsc{ii} ]\lambda 6731/6717 \else [S~{\scshape  ii}]$\lambda 6731/6717$\fi\xspace}

\newcommand{\Toiii}{$T_{\rm e}$\oiii\xspace}
\newcommand{\Tnii}{$T_{\rm e}$\nii\xspace}

\newcommand{\Ho}{\ensuremath{\mathrm{H}^{0}}\xspace}
\newcommand{\Hp}{\ensuremath{\mathrm{H}^{+}}\xspace}

\newcommand{\Hep}{\ensuremath{\mathrm{He}^{+}}\xspace}
\newcommand{\Hepp}{\ensuremath{\mathrm{He}^{++}}\xspace}
\newcommand{\Op}{O$^{+}$\xspace}
\newcommand{\Opp}{O$^{++}$\xspace}

\newcommand{\woiii}{\ifmmode \rm{W[O}\,\textsc{iii}] \else W[O~{\sc iii}]\fi\xspace}
\newcommand{\wha}{\ifmmode {W(\rm h}\alpha )\else W(H$\alpha$)\fi\xspace}

\begin{document}

\lefttitle{G. Stasi\'nska}
\righttitle{Determining the chemical composition of planetary nebulae}

\journaltitle{Planetary Nebulae: a Universal Toolbox in the Era of Precision Astrophysics}
\jnlDoiYr{2023}
\doival{10.1017/xxxxx}
\volno{384}

\aopheadtitle{Proceedings IAU Symposium}
\editors{O. De Marco, A. Zijlstra, R. Szczerba, eds.}
 
\title{On determining the chemical composition of planetary nebulae
}

\author{Gra\.zyna Stasi\'nska}
\affiliation{LUTH, Observatoire de Paris, CNRS, Universit\'e Paris Diderot; Place Jules Janssen, F-92190 Meudon, France (ORCID  0000-0002-4051-6146, e-mail: grazyna.stasinska@obspm.fr)}

\begin{abstract}

We present literature on abundance determinations in planetary nebulae (PN) as well as public tools that can be used to derive them. Concerning direct methods to derive abundances we discuss in some depth such issues as reddening correction, use of proper densities and temperatures to compute the abundances, correction for unseen ionic stages, effect of stellar absorption on  nebular spectra, and error analysis. Concerning photoionization model-fitting, we discuss the necessary ingredients of model stellar atmospheres, the problem of incomplete slit covering and the determination of the goodness of fit. A note on the use of IFU observations is given. The still unsolved problem of temperature fluctuations is briefly presented, with references to more detailed papers. The problem of abundance discrepancies is touched upon with reference to more extensive discussions in the present volume.
Finally carbon footprint issues are mentioned in the context of extensive PN modeling and large databases.

\end{abstract}


\maketitle

\section{Introduction}
\label{basic}

The first elemental abundance determinations in planetary nebulae (PN) date back to the study of Bowen \& Wyse (1939), who elaborated a way to derive the abundances from the observed emission lines and showed that, in spite of PN spectra being very different from those of stars, their chemical composition is very similar.

Since then, abundances studies in PNe have flourished, with the aim both to determine the chemical composition of the gas out of which they were formed and to estimate the amount of element production by the PN progenitors. 

For any PN, the derived chemical composition depends on the available observations, on the adopted methods and on the atomic data that have been used. Table \ref{tab1} shows the abundances of He, C, N, O, Ne and S determined by various authors for the brightest PN: NGC 7027, using various methods. The bottom line shows the standard deviation of all these determinations, and gives a \textit{rough }idea of the expected uncertainties in PN abundances including all possible sources of errors. 

\begin{table} [h]
  \begin{center}
  \caption{Comparison of abundances for NGC 7027 in units of log X/H + 12.}
  \label{tab1}
  \begin{tabular}{|l|c|c|c|c|c|c|}
  \hline 
							& He 	& 	C 	&	N   &	O  &  Ne  & S	 \\
\hline
Stanghellini et al. 2006	& 11.21	&       & 8.43	& 8.43 & 7.67 &      \\
Zhang et al. 2005		    & 11.00 & 9.10	& 8.14	& 8.66 & 8.07 & 6.92 \\
Kwitter et al. 2003		    & 11.00	&       & 8.62	& 8.37 & 7.85 & 6.64 \\
Bernard-Salas et al. 2001	& 11.03	& 8.78	& 8.20	& 8.61 & 8.00 &	6.97 \\
Kwitter \& Henry 1996		& 11.00	& 8.98	& 8.21	& 8.71 & 8.14 & 7.85 \\
Keyes et al. 1990		    & 11.05	& 8.84	& 8.10	& 8.49 & 8.00 & 6.86 \\
Middlemass 1990		        & 11.02	& 9.11	& 8.28	& 8.75 & 8.04 & 6.90 \\
Perinotto et al. 1980		&       & 9.11	& 8.52	& 8.62 &      &      \\
P\'equignot et al. 1978		& 11.00	& 9.48	& 8.30	& 8.86 & 8.34 & 7.23 \\
Aller 1954	        	    & 10.96	&       & 8.52	& 8.96 & 8.44 & 8.17 \\
\hline
Standard deviation			& 0.06   & 0.21	& 0.17	& 0.18 & 0.22 & 0.50 \\
\hline
  \end{tabular}
 \end{center}
\end{table}

A recent study by Rodr\'iguez (2020) compared abundances derived with the same methods and same atomic data for about 20 Galactic PNe that have more than one spectrum of good quality. The derived one-sigma observational uncertainties from these objects are: 0.11 dex for O/H, 0.14 dex for N/H, 0.14 dex for Ne/H, 0.16 dex for S/H, 0.11 dex for Ar/H, and 0.14 dex for Cl/H an less than 0.05 dex for He/H.

Several reviews have been published recently on abundance determinations in ionized nebulae with special attention to PNe, (eg, Stasi\'nska  2004, 2009, Peimbert et al. 2017). The present text is not a tutorial on abundance determination, but a complement to the afore mentioned papers incorporating recent studies and emphasizing issues that need a careful treatment.

There are basically two ways to determine element abundances in PNe from their emission lines. One is the `direct method' which obtains ionic abundances from ratios of the observed emission-line fluxes and next derives total elemental abundances by using ionization-correction factors (ICFs). The other is to fit the observed line intensities using photoionization models. 
Both approaches will be discussed here, but first we will examine the issue of correcting for extinction due to interstellar dust.


\section{Reddening correction}
\label{redd}

\subsection{Generalities}

The nebular flux after attenuation by a purely absorbing intervening dust slab  is given by:
\begin{equation}
\label{eq.red1}
F(\lambda)_{\rm{obs}} = F(\lambda)_{\rm{em}} \times \rm{exp}(-\tau(\lambda)),
\end{equation}
where $F(\lambda)_{\rm{em}}$ is the emitted flux at wavelength $\lambda$ and $F(\lambda)_{\rm{obs}}$ is the observed flux at this wavelength.

In nebular astrophysics, one often uses the following expression to relate $F(\lambda)_{obs}$ and $F(\lambda)_{em}$:
\begin{equation}
\label{eq.red2}
F(\lambda)_{\rm{obs}} = F(\lambda)_{\rm{em}} \times 10 ^{C f(\lambda)}
\end{equation}

where $C$ is the logarithmic extinction at \Hb and $f(\lambda)$ is called the `extinction law' considered as universal, equal to 1 at  4861\AA.

Knowing the form of  $f(\lambda)$ , $C$ can be derived from the observed \Ha/\Hb ratio by comparing it to the theoretical one for case B recombination:
\begin{equation}
\label{eq.red3}
C = [ \rm{log}(F(\Ha)/F(\Hb)_{\rm{B}} - \rm{log}(F(\Ha)/F(\Hb)_{\rm{obs}} ] / (f(\Ha) -f(\Hb))
\end{equation}

Emission line ratios are then dereddened using the formula:
\begin{equation}
\label{eq.red4}
\rm{log} (F(\lambda_1)/F(\lambda_2))_{\rm{corr}} =  \rm{log} (F(\lambda_1)/F(\lambda_2))_{\rm{obs}}  + C \times (f(\lambda_1) -f(\lambda_2))
\end{equation}

\subsection{Complications}

\begin{itemize}
 \item  The intrinsic \Ha/\Hb ratio depends slightly on the electron temperature \Te\ and electron density \Ne.
This can easily be dealt with a iterative scheme,  \ie first dereddening the line ratios using  canonical values for \Te\ and \Ne\ (for example 10,000 K and 100  \cmcub), then computing \Te\ and \Ne\ from dereddened line ratios and iterate, as advocated \eg by Ueta \& Otsuka (2021). Such a procedure has been adopted in many studies. However, unless the temperature or densities are very different from the ones that were chosen a priori, the error on abundances is not very large (only of a few percent for 8,000 K $<$ \Te $<$  12,000K, Morisset et al. 2023) so the results obtained by authors who, for some reason, do not apply an iterative approach, remain acceptable. 
  \item  The intrinsic \Ha/\Hb ratio may be different from that of the theoretical case B due to collisional excitation of \Ho. The effect is larger in objects where the proportion of neutral hydrogen is larger and the electron temperature is high. For example for \Te  equal for 17,000 K and  $\Ho/H = 10^{-2}$,  \Ha/\Hb is equal to 3, while it would be of 2.75 in absence of collisional excitation (see Fig. 1 of Luridiana 2009).
  \item  The extinction law is actually not universal. It has been shown by Cardelli et al. (1989) that its form can be parametrized by the parameter: $R_V = A(V )/E(B-V)$,  \ie the ratio of total to selective extinction at $V$. While 3.1 is the standard value adopted for $R_V$ in the vast majority of studies, the actual values vary from 2.3 to 5.5 (Cardelli et al. 1989, Fitzpatrick 1999). This is almost never considered in spectroscopic studies of PNe. The value of $R_V$ is related to the environment. Large values of $R_V$, which correspond to a flatter extinction law, are found for lines-of-sight crossing dense interstellar clouds, where grain growth occurs, while lower values of  $R_V$ are found in the diffuse interstellar medium. In absence of any information on the value of $R_V$, the uncertainties on reddening correction due to the form of the `extinction law' can be taken into account using Table 1 of Fitzpatrick (1999). 
  \item If some dust is mixed with the ionized gas and strongly contributes to the extinction, no `extinction law' applies. Such a situation, however, is unlikely to occur in planetary nebulae, except in very young and compact objects. 
  \item Instead of using only \Ha and \Hb one can force the observed Balmer decrement to its theoretical value.
Such a procedure avoids the problem of the extinction law 
as well as of the flux calibration in the blue (which is difficult to achieve)  
but it requires a correct treatment of the observational uncertainties.
This procedure returns correct line ratios even in the case where internal dust plays a role. However, is feasible only if the signal-to-noise ratio of all the lines that are used is sufficiently high.
  \item  In many cases only \Ha and \Hb intensities are of sufficient accuracy so that the use of an extinction law cannot be avoided.
In case of calibration problems, the resulting intensities of  H$\gamma$, H$\delta$ etc.  may be far from the theoretical values. 
This has to be artificially corrected, otherwise the intensities of nearby lines will be wrong. An important case is that of the \Oiiit line, which lies near the  H$\gamma$ line, and has a strong impact of the derived chemical composition of the PNe.
  \item The observed Balmer lines can be affected by stellar absorption. This happens when the slit covers the central star.  The effect of absorption on Balmer lines can be tackled with an adjusted model of the stellar atmosphere or by observations with good spectral resolution 
(stellar lines are generally broader than emission lines). An example of such a procedure is described in Sect 2.2.3 of Stasi\'nska et al. (2010). Another case is that of extragalactic PNe, when the slit encompasses a lot of light from the stars located in the galaxy. The best practice to deal with such a case would be to fit the observed stellar continuum using a stellar population synthesis code and subtract it from the observed spectrum. Such a procedure is routinely applied in studies of emission-line galaxies (see \eg Asari et al. 2007, Thomas et al. 2013, Belfiore et al. 2022), and should also be implemented in studies of extragalactic PNe. 
\end{itemize}

\section{Abundance determinations using direct methods}
\label{direct}

Direct methods are very simple in their principle. The abundance ratio of two ions is obtained from the observed intensity 
ratio of lines emitted by these ions.  For example,  
\Opp/\Hp\ can be derived from 
\begin{equation}
{\rm O ^{++}/ H^{+}} = \frac{{\rm[ O \,{\sc{III}}]}\,\lambda 5007/{\rm 
H}\beta} 
{j_{{\rm [ O \,{\sc III}]}(T_{e},n_{e})}/j_{{\rm H}\beta (T_{e},n_{e})}}.
\end{equation}
$j_{\rm{[O}\,{\sc III}]}(T_{e},n_{e})$ and  $j_{\rm H}\beta (T_{e},n_{e})$ are the emission coefficients of the 
\Oiii\ and \Hb lines respectively, which depends on \Te\ and \Ne (assumed uniform in the nebula). Both \Te and \Ne can be derived from specific line ratios.

However, as will be seen, the devil is in the details.		

\subsection{Softwares for direct abundance determinations}
\label{softwares}

Over the years, many softwares have been developed for public use: ABELION (Stasi\'nska 1978), FIVEL (de Robertis et al. 1987), NEBULAR (Shaw \& Dufour 1995), ELSA (Johnson et al. 2006), NEAT (Wesson et al. 2012), PyNeb (Luridiana et al. 2015). 
Of all these softwares, PyNeb (https://pypi.org/project/PyNeb/) is the most complete and versatile, regularly updated by C. Morisset and deserves a more detailed description here.  

PyNeb is easy to install.  It is written in Python and uses standard Python libraries such as numpy, matplotlib, pyfits, scipy and other. 
While PyNeb was originally designed to compute nebular abundances, it can perform many tasks which help understand the processes giving rise to the intensities of emission lines in ionized nebulae and thus have a better control on the results. Examples of things PyNeb can do: 
 
\begin{itemize}
  \item compute physical conditions from suitable diagnostic line ratios.
  \item compute level populations, critical densities and line emissivities.
  \item compute and display emissivity grids as a function of \Te and \Ne.
  \item correct line intensities for reddening using various extinction laws.
  \item read and manage observational data.
  \item plot and compare atomic data from different publications.
  \item compute ionic abundances from line intensities and physical conditions.
  \item compute elemental abundances from ionic abundances and ICFs. 
\end{itemize}

Note however that, when using PyNeb \textit{it is the user’s responsibility } to make all the choices concerning reddening correction, electron temperatures and densities, etc. These choices must be clearly described in the publications. 
Also, it is important to always mention the version of PyNeb and list the references for the atomic data that are being used.
Before publishing, it is recommended to analyze the reliability of the results and perform some checks by computing a few abundances `by hand' using the emission-line tables that can be produced by PyNeb. 

\subsection{Choosing a set of atomic data}
\label{atdat}

As shown by Juan de Dios \&  Rodr{\'\i}guez (2017), atomic data variations introduce differences in the derived abundance ratios as low as $0.1–0.2$ dex at low density, but that reach $0.6–0.8$ dex at densities above $10^4$ cm$^{-3}$ in several abundance ratios, such as O/H and N/O. 

Note that the most recent atomic data are not always the best. Even if the methods are more refined, computations may still involve approximations or even errors. 
For example, Palay et al. (2012) were the first to calculate collision strengths for the O III forbidden transitions using a relativistic Breit–Pauli R-matrix method with resolved resonance structures. Their results were significantly different from those of previous authors. However, Storey et al. (2014) noted that Palay et al. (2012) had omitted the three 2p$^4$ terms from their scattering target, leading to a downshift of the temperature derived from \rOiii by up to 600 K. 

A default data set is now provided by PyNeb thanks to the collaboration with C. Mendoza, and will be regularly updated (Morisset et al. 2020, Mendoza et al. 2023). In any case, it is always necessary to publish the exact source of atomic data that are used in a given computation.

\subsection{What about the general density stucture of PNe?}
\label{densitystructure}

The density derived from \rAriv is generally larger than that derived from \rSii (e.g. Wang \& Liu 2007, M{\'e}ndez-Delgado et al. 2023). 
This can be due either to the different sensitivity of the line ratios or to density stratification in PNe.  If \rSii is not available, or very uncertain, this has an important consequence on the derived N/O abundance ratio, which could be overestimated by up to 50\%. 

However, comparing the densities derived from \rSii, \rOii, \rCliii and \rAriv in a large sample of PNe Juan de Dios \&  Rodr{\'\i}guez 2021  concluded that `the density structures derived for the objects depend strongly on the atomic data used in the calculations: atomic data are shaping the derived density pattern.'

Another and completely independent way to derive the density structure of PNe is to use \Ha surface brightness  maps  (see \eg Stasi\'nska et al. 2010). However, no systematic study comparing this approach to that of line ratios has been made yet.

All the above concerns large scales. But what happens when looking on smaller scales? The James Webb Space Telescope has shown that the bright shell of the ring nebula,  NGC 6720, is fragmented in thousands of dense globules of molecular gas, with a characteristic diameter of 0.2 arcsec and density of $10^5-10^6$ cm$^{-3}$, embedded in ionized gas (Wesson et al. 2023). This implies that the ionized gas at the frontier of these globules cannot be of uniform density.  Chemical abundances derived under the hypothesis of constant density are therefore likely to be biased.
\subsection{Choosing \Te and \Ne}
\label{tene}

In direct abundance determinations in PNe, the common practice is to build a \Te, \Ne diagram for the various observed line ratios that are sensitive to \Te and/or \Ne. From such a diagram one reads out the values of \Te and \Ne representative of the different zones in the nebula. 

This empirical procedure needs some clarifications. First of all, it is important to visualize the effect of the error bars on the line ratios. This is easily achieved with the PyNeb software. Even so, the various temperature or density indicators do not always seem to give reasonable values of \Te and \Ne (see \eg Garc\'ia-Rojas et al. 2012). However, one must recall that these diagrams are built under the hypothesis of uniform \Te and \Ne for each diagnostic line ratio, which obviously cannot be the case in real objects. So any choice of \Te and \Ne from such diagrams is just an educated guess which  gives an incomplete (and biased) idea of temperature and density stratifications. Abundances derived from different lines of the same ion may differ. For example, in NGC 7027, adopting \Te =  12,600 K for all the lines emitted by \Opp 
 Zhang et al. (2005) find  \Opp/\Hp $= 2.56 10^{-4}$ from the $\text{O}\,\textsc{iii}]\lambda 1661$ line  but \Opp/\Hp $=3.08 10^{-4}$ from the \Oiii and \Oiiit lines. This difference is due to the fact that ultra-violet lines are more sensitive to high temperatures. 

\subsection{More on temperature diagnostics}
\label{tenemore}

As known from some time, collisionally excited lines (CELs) can be affected by recombination of the upper ion. This is the case, for example, of  $[\text{N}\,\textsc{ii}]\lambda 5755$,  $[\text{O}\,\textsc{ii}]\lambda 7720+7730$  and  $[\text{O}\,\textsc{iii}]\lambda 4363$, which are all used for temperature diagnostics. 
Liu et al. (2000) proposed formulae to estimate the recombination contribution. These formulae depend on the  abundance of the upper ion and on \Te. They are generally used assuming that \Te is equal to the temperature derived from \rOiii. But this is wrong in the presence of cold zones, and the actual recombination contribution is larger than computed!
G\'omez-Llanos et al. (2020) proposed a formula to correct \Oiiit for recombination for the case if high excitation PNe, but they do warn against its blind use  in cases where the gas has two phases of different chemical composition, with the recombination contribution coming mainly from the cold regions.

Garc\'ia-Rojas et al. (2022) have proposed a more direct empirical method using the strongest observed recombination lines  $\text{N}\,\textsc{ii}\lambda 5679$ and $\text{O}\,\textsc{ii}\lambda$ 4649+50, whose emissivity ratio is almost independent of \Te. To correct  the $[\text{N}\,\textsc{ii}]\lambda 5755$ for the contribution of recombination they write:

\vspace{-\baselineskip}
\begin{equation}
\label{rec}
I(\lambda 5755)_{\rm corr} =I(\lambda 5755) - j(\lambda 5755)/j(\lambda 5679) \times I(\lambda 5679) 
\end{equation}
It must be added that recently Nemer at al (2019) reported the first observational evidence of Rydberg Enhanced Recombination (RER), an important recombination process that has so far been unexplored. First estimates on carbon ions shows that RER has a significant impact on the ionization structure of ionized nebulae containing cold zones and on the derived abundances.

\subsection{Correction for unseen ionic stages}
\label{icf}

The ICFs computed by Kingsburgh and Barlow (1994) on the basis of a few tailored PN models have been used for decades. 

In 2014, Delgado Inglada et al. have published ICFs for He, O, N, Ne, S, Ar, Cl, and C, based on a grid of ionization- and density-bounded photoionization models of different values of stellar luminosities and effective temperatures, nebular masses and internal radii. 
The models were computed with the photoionization code CLOUDY c10.01 (Ferland et al. 1998). The initial grid was trimmed so as to keep only those models that could represent PNe in terms of nebular masses, surface brightness and exciting star masses. This resulted in 2820 models which were used to compute analytical expressions of ICFs for different ions as a function of \Opp/(\Op+ \Opp) or \Hepp/(\Hep+ \Hepp), which are quantities that can be obtained from the observations. With this procedure, the authors were able to estimate for the first time the values of the uncertainties linked to ICFs. However, it turns out that the grid of models was slightly too large and that the uncertainties were not well evaluated. More recently, Amayo et al. (2021) produced ICFs for giant \hii regions, based on a much more refined grid of models taking into account the probability that each model can represent a giant \hii region in the real world and using a more correct procedure to estimate the uncertainties. Such an enterprise should also be undertaken for PNe. 

Another option is now offered by machine learning techniques, which can easily provide ICFs that are tailored to a given object. For example, Sabin et al. (2022) and Garc\'ia-Rojas et al. (2022) have used an enhanced grid of PN photoionization models (3MdB, Morisset et al. 2015,  https://sites.google.com/site/mexicanmillionmodels/) 
and selected the models that have \textit{all} of the He$^{++}$/He$^{+}$, O$^{++}$/O$^{+}$,  Ne$^{4+}$/Ne$^{3+}$, Ne$^{3+}$/ Ne$^{++}$, and Ar$^{4+}$/Ar$^{3+}$/ ratios close to the observed ones. With such a technique, it should also be easy to estimate error bars.

\subsection{Error analysis}
\label{error}

The best way to perform reliable error analyses is through  Monte-Carlo techniques.
A proper analysis should start right from data acquisition 
and go through all the processes involved in abundance determination, that is: 

\begin{itemize}
  \item Photometric calibration.
  \item Line intensity measurement.
  \item Extinction `law'.
  \item Reddening computation
  \item Determination of plasma parameters \Te and \Ne
  \item Ionization correction factors
\end{itemize}

\smallskip

Some difficulties are however involved in the process, which are not always recognized.
\begin{itemize}
  \item The form of the distribution in errors in line intensities is not always known, and depends on the detection and measurement process:  gaussian, log gaussian, something else? (See Wesson et al. 2016).
  \item If only one temperature diagnostic is available, \eg \Toiii, one has to adopt a relation between its value and the values of the other characteristic temperatures, \eg \Tnii. What distribution to use for this relation?
  \item What distribution to consider for uncertainties in the ICFs?
\end{itemize}

\smallskip

It is not sufficient to state that uncertainties have been obtained with Monte-Carlo technique. A proper description is necessary. For example in the case of gaussian distributions, the intensities may become negative. How is this taken into account? When using several hydrogen lines to determine the reddening and simultaneously determine the plasma parameters \Te and \Ne, it is important to first determine which hydrogen lines are to be included in the process, and to check the final results for reliability (for example the final abundance results of Ueta \& Otsuka 2022 on the N/O abundances are obviously wrong since in some of the PNe they studied, the found a `corrected'  \Ha/\Hb ratio that is impossible to achieve in any nebula (see Morisset et al. 2023 for a detailed discussion of this case).

Finally let us point out that resulting abundance distributions may be strongly asymmetric so it is recommended to use medians and quartiles rather than means and dispersions to describe them.

\section{Abundance determination with photoionization models}
\label{photo}

Many studies have considered that, especially in absence of direct information on the electron temperature, \eg \Toiii, the chemical composition of PNe can be derived through photoionization model fitting. This is only partly true. First of all, in absence of adequate constraints, the solution may not be unique, while generally the search is stopped after one solution is found. A proper photoionization modeling may give incorrect abundances if the nebula is density bounded while the model is ionization bounded, or if helium lines are not fitted.

The most recent review dealing with photoionization models is the one by Morisset (2017). Here we mainly comment on the question of abundance determination using photoionization models.

Many photoionization codes have being written, but only a few are of public use. They are listed below.

\subsection{Open-source photoionization codes}
\label{open}
 

\subsubsection{Spherical Geometry}
\textbf{Cloudy.} This photoionisation code has been in continuous development since 1978, led by Gary Ferland, and
in collaboration with many scientists. The latest full description in a journal is  by Ferland et al. (2017) and the website https://gitlab.nublado.org/cloudy/cloudy. 
The current version is C23. 
This code is the most widely used to model dusty \hii regions, PNe and active galactic nuclei. It comes along with a thorough documentation and contains a very complete and regularly updated collection of atomic processes including molecules.

\textbf{Mappings V. } The original MAPPINGS code (Binette et al. 1985) was developed to model the emission line and continuum spectra of  \hii regions, PNe and active galactic nuclei and do the same for plasmas that are out of collisional or photoionization equilibrium, such as the radiative shocks in supernova remnants and their precursor zones. The interaction with dust was introduced in 2000. Mappings V  has an expanded database for cooling and recombination lines and is described in Sutherland \& Dopita (2017). It can be found on https://bitbucket.org/RalphSutherland/mappings/src/public/

\vspace{-\baselineskip}

\subsubsection{Pseudo-3D} 
\textbf{pyCloudy.} This is a Python library developed by Morisset (2013) to generate 3D nebula models from various runs of the 1D Cloudy code. It can also be used more generally to handle Cloudy input and output files and easily build grids of Cloudy models. It has been employed to build the models of the 3MdB database. 
It is found at https://pythonhosted.org/pyCloudy/. Note that, although the treatment of the diffuse ionizing radiation is largely simplified, pseudo-3D models are a convenient substitute to full 3D models in the case where the PN geometries are not too complicated and have the advantage of being much faster, which is appreciable when carrying-out model-fitting of PNe observations (\eg Gesicki et al. 2016)

\vspace{-\baselineskip}
\subsubsection{3D}
\textbf{Mocassin.} This is a fully 3D or 2D photoionization and dust radiative transfer code which uses a Monte Carlo approach to the transfer of radiation through media of arbitrary geometry and density distribution (Ercolano et al.  2003, 2005, 2008) https://mocassin.nebulousresearch.org/. 
It allows modeling PNe with such extreme morphologies as NGC 6302 (Wright et al. 2011).

\textbf{Messenger Monte Carlo MAPPINGS V (M$^3$)}   
M$^3$ (Jin et al. 2022) is a  3D photoionization code, combining the Monte Carlo radiative transfer technique with the MAPPINGS V photoionization code. Its purpose is to produce reliable diagnostic emission lines in ionized nebulae with arbitrary geometry. This code is not yet publicly available.

\subsection{Why and how use photoionization models for abundances?}
\label{fitting}

Photoionization model-fitting is much more time-consuming than direct abundance determination. So, is it worthwhile undertaking such a task?
When no direct determination of \Te are available, this is the only way to put some limits on the abundances. Even  if \Te measurements exist, it may be useful to consider photoionization modeling, because there is no ICF issue: the chemical composition of the object under study is that of the model which fits all the lines.

The model is supposed to take into account all the relevant physical processes occurring in the nebula. The chemical composition of the nebula is that of the model that fits \textit{all} the observed line ratios. 

The model should use properly all the available constraints
\begin{itemize}
  \item density structure from an \Ha image,
  \item line ratios (in the appropriate apertures),
  \item properties of the exciting star, if known,
  \item luminosity, if known.
\end{itemize}

While often published models of PNe assume that they are  ionization-bounded, this is not necessarily the case. They can be matter-bounded, especially in some directions.

If the observational slits do not cover the entire nebula, the comparisons with the models must account for it, by extracting from the model a portion of the nebula which corresponds to the slit, like \eg in Morisset \& Georgiev (2009), or Stasi\'nska et al. (2010).

One important ingredient of photoionization modeling is the description of the ionizing radiation. While blackbodies can be useful for exploratory studies, proper model atmospheres of the central stars are required for a detailed model-fitting. Such models are discussed by Kudritzki et al. (2006). Grids of plane-parallel NLTE model atmospheres of hot white dwarfs using TMAP have been computed by Rauch (2003), Rauch \& Reindl (2014). NLTE models for central stars with winds are computed by using CMFGEN (Hillier 2006) and using METUJE by Krti{\v{c}}ka et al. (2020). 

A model should fit \textit{all} the observed line ratios within the uncertainties (observational and model).
If it does not (which is often the case), it is not easy to determine what are the real abundances. Perhaps some important physical parameter is not correctly treated or missing (erroneous assumed geometry, distribution of dust grains, missing of cooling or heating mechanisms etc.)

To estimate the goodness of a model, it is not recommended to use a khi-square approach as is sometimes done (\eg Henry et al. 2015) because \textit{each }of the mismatched line intensity has a significance that is worth investigating. One can instead (Morisset \& Georgiev 2009, Stasi\'nska et al. 2010, Bandyopadhyay et al. 2021, Miranda-Marques et al. 2023) use a quality factor for each observable  
\begin{equation}
\label{ }
\kappa(\rm{O}) = (\rm{log} O_{\rm{mod}} - \rm{log} O_{\rm{obs}})/\tau(O)
\end{equation}
where $\tau$(O) is the accepted tolerance in dex, which takes 
 into account the observational uncertainty in flux ratio and reddening as well the expected ability of the model to reproduce the observable.
  $\tau$(O)  is defined as follows: 
\begin{equation}
\label{ }
\tau(\rm{O}) = \rm{log} (1 + \Delta (O)/O),
\end{equation}
where $\Delta$(O) is the absolute value of the maximum `acceptable' error on the observable, 
A good model should have all the values of  $\kappa$(O) between $-1$ and $+1$.
Of course, a preliminary condition is that the model returns the correct value of the \Ha flux.

How to estimate error bars on elemental abundances? They cannot be easily deduced from errors in the observed line intensities.
One must find \textit{all} the models that are compatible with the observational error bars. This would be a huge task, almost never undertaken.
Stasi\'nska et al. (2010) have done such an exercise for the most oxygen-poor PN, PN G 135.9+55.9, but this was a relatively easy case, because the abundances of metals being small, they did not impact on the thermal balance. 

There is a remaining problem with abundances derived from models. It often occurs that, while the intensities of the strongest lines are fitted, those of the weak lines are not, in particular those which indicate the value electron temperature (see \eg Kwitter \& Henry 1996, 1998). Such a mismatch indicates that the temperature of the model is incorrect and therefore the chemical composition of the model cannot be considered as the correct one. Such cases require a discussion to understand why the predicted thermal balance is off.

\subsection{Do integral field unit (IFU) observations help for abundance determinations?}
\label{ifu}

Spectra provided by IFUs are extremely valuable in PNe studies since they allow one to determine the distributions  of
\begin{itemize}
  \item extinction,
  \item densities,
  \item local inhomogeneities (clumps, filaments, low ionization structures),
  \item electron temperatures,
  \item zones possibly affected by shocks,
  \item ionic ratios,
  \item ionic abundance discrepancy factors (see next section).
\end{itemize}

But they do not allow a direct determination of the distribution of elemental abundances, because the ICFs on different lines of sight are not known. The available ICFs have been obtained using PN photoionization models that represent entire objects and cannot be applied on a spaxel-by-spaxel basis. However, adding up all the spaxels in a given object provides an integrated spectrum which, indeed, can be treated by the direct method.

The integrated spectrum, together with the observed surface brightness distribution, is also ideal to confront with self-consistent photoionization models, which are, at the same time, required to reproduce the observed distribution of electron temperatures and densities (see Basurah et al. 2016). Undertaking this with all the accuracy allowed by the observations is a phenomenal task successfully achieved by H.Monteiro (these proceedings).

\section{Temperature fluctuations and abundance discrepancies}
\label{tfluct}

Before leaving, we must evoke a very important problem which has been with us for decades and is still harshly debated. It is the so-called problem of `temperature fluctuations' and the associated one of 'abundance discrepancies'. Temperature fluctuations were suggested by Peimbert (1967) to explain the discrepancy between electron temperatures derived from collisionally excited lines and those derived from the hydrogen Balmer jump. A technique was then proposed to correct the derived abundances for this effect. These temperature fluctuations are often ignored in direct abundance derivations. The question of the discrepancy between abundances derived from collisionally excited lines and from recombination lines is thoroughly discussed by  M{\'e}ndez-Delgado \& Garcia-Rojas and by Morisset et al. (these proceedings).
Previous reviews on these topics are by Esteban (2002), Liu (2002), Ferland (2003), Peimbert \& Peimbert (2003), Stasi\'nska (2004, 2009), Garc{\'\i}a-Rojas  et al. (2019), where many references can be found.

Some recent studies have provided an important progress in the observational evidence of the existence of temperature fluctuations (Peimbert \& Peimbert 2013, M{\'e}ndez-Delgado et al., 2023b) since now the evidence is based on lines from \Opp only and does not require additional information from hydrogen lines which introduced some approximation in the derivation. However, it has not been proven that the formulation of Peimbert (1967) to derive the temperature fluctuation parameter $t^2$ gives correct results in the case of extreme temperature jumps such as may occur even in the case of a canonical $t^2= 0.04$ (see Stasi\'nska 2004). Indeed the formulation is based on Taylor series which may not be appropriate in this case. The other remaining problem is to find mechanisms that can \textit{quantitatively} reproduce the inferred `fluctuations'. Perhaps the more promising one is photoelectric heating of small grains in nebulae with density condensations which, as showed by Stasi\'nska \& Szczerba (2001) spectacularly boosts the temperature of the diluted component. However, there is presently no direct evidence of the presence of small grains in such nebulae.

Regarding abundance discrepancies, the presence of oxygen-rich inclusions in some PNe has also been clearly demonstrated (see Espiritu \& Peimbert 2021 and references in M{\'e}ndez-Delgado \& Garcia-Rojas, these proceedings) as well as their concentration in the central zones. It has also been shown  that the most extreme abundance discrepancies are found in PNe with binary central stars (Wesson et al. 2018) suggesting than they may be caused by a nova-like eruption occurring soon after the common-envelope phase.

\section{Carbon footprint issues}
\label{carbon}

While the complexity of computations in the field of PNe is far from comparing with that of cosmological or weather forecast simulations in terms of both CPU and memory, it is still worth asking the question of the carbon footprint of computational tasks (Lannelongue et al. (2001).

With modern computers, calculating an abundance or producing a photoionization model is almost instantaneous. To the point that not much effort is spent nowadays on reducing computing times (in contrast to the situation in the 70-80'ties when computations were much slower and optimization was vital).

However, with IFUs now producing spectra in many spaxels (\eg 40,000 spaxels for MUSE observations of PNe), with Monte-Carlo simulations requiring random sampling for each process and each spaxel, computation times may become huge.



For example, for the 724,386 PN entries in the 3Mdb\_17 database which required 145,150 runs of Cloudy 17.01, the total computation time would be of nearly 500 days if using one processor on an Intel(R) Xeon(R) CPU E5-2640-v3@2.60GHz. With 50 processors, the time is reduced by a factor 50. Note that if the code had been paralleled (which is not the case here), the carbon footprint would at the same time have increased by a factor of 2 (see Fig. 3 in Lannelongue et al 2021).

Therefore, reducing computing times again becomes a concern nowadays. The first thing is to improve computer programs with efficient coding, and, perhaps, to use more suitable programming languages. For the building of large grids of PNe models, an alternative to Cloudy would be needed since Cloudy, because it treats so many physical processes in detail, is very slow compared to other photoionization codes that are no so complete but still useful for most studies on PNe (\eg PHOTO, last described in Stasi\'nska 2005 is at least 10 times faster).

Another road is to use databases of precomputed models (\eg 3MdB) as much as possible. 
Finally it is a good idea to ask oneself whether the gain in scientific result is worth the computational cost. Artificial intelligence (AI) techniques are now very useful to reduce the number of simulations in a given project. Some
examples can be found in Sabin et al. (2021) or Garc\'ia-Rojas et al. (2022).

\section{Summary}
\label{take}

Over the 80 years of abundance determinations in PNe, the basic methods have remained the same. Abundances are believed more reliable today mainly due to progress in atomic data computations. PNe in external galaxies are now within reach for abundance analyses thanks to large telescopes and appropriate instrumentation. Public tools are now available for direct abundance derivation as well as for photoionization modeling (but should not be used without caution).

Here we have revised some problems which still need attention such as reddening correction, use of proper densities and temperatures to compute the abundances, correction for unseen ionic stages and error analysis.

As in many fields in astrophysics, machine learning techniques are starting to be used to perform specific tasks such as deriving ionization correction factors based a grids of photoionization models or speeding up computational process in Monte-Carlo estimation of plasma parameters in the case of IFU observations of PNe.

Contrary to what one might believe, elemental abundances derived from photoionization modeling are not necessarily better than abundances derived from a direct analysis of the spectra and the related uncertainties are very difficult to assess.

Important progress in the long standing problem of temperature fluctuations and abundance discrepancies (see Sect. \ref{tfluct}) has been made, but the topic remains the most challenging one in nebular astrophysics.


\end{document}